\pdfoutput=1
\documentclass[pra,reprint,showpacs,amsmath,amssymb]{revtex4-1}
\usepackage{hyperref}
\usepackage{graphicx}
\usepackage{xspace,xcolor,textcomp}
\usepackage[figure,table]{hypcap}

\definecolor{darkblue}{rgb}{0.459, 0.439, 0.702}
\hypersetup{colorlinks=true,
linkcolor=blue,
filecolor=blue,
citecolor=blue,
urlcolor=blue,
pdfauthor={Matthias Rosenkranz, Yongyong Cai, and Weizhu Bao},
pdftitle={Effective dipole-dipole interactions in multilayered
  dipolar Bose-Einstein condensates},
pdfkeywords={dipolar BEC, optical lattice}}

\newcommand{\eg}{\textit{e.g.}\xspace}

\newcommand{\bv}[1]{\ensuremath{\boldsymbol{\mathrm{#1}}}}
\newcommand{\im}{\ensuremath{\mathrm{i}}}
\newcommand{\eu}{\ensuremath{\mathrm{e}}}

\DeclareMathOperator{\erfc}{erfc}
\DeclareMathOperator{\erf}{erf}
\DeclareMathOperator{\sgn}{sgn}
\DeclareFontFamily{U}{euc}{}
\DeclareFontShape{U}{euc}{m}{n}{<-6>eurm5<6-8>eurm7<8->eurm10}{}%
\DeclareSymbolFont{AMSc}{U}{euc}{m}{n} 
\DeclareMathSymbol{\umu}{\mathord}{AMSc}{"16}

\begin{document}
\title{Effective dipole-dipole interactions in multilayered dipolar Bose-Einstein condensates}
\author{Matthias Rosenkranz}
\author{Yongyong Cai}
\author{Weizhu Bao}
\affiliation{Department of Mathematics, National University of
  Singapore, 119076, Singapore}
\date{\today}

\begin{abstract}
  We propose a two-dimensional model for a multilayer stack of dipolar
  Bose-Einstein condensates formed by a strong optical lattice.  We
  derive effective intra- and interlayer dipole-dipole interaction
  potentials and provide simple analytical approximations for a given
  number of lattice sites at arbitrary polarization.  We find that the
  interlayer dipole-dipole interaction changes the transverse aspect
  ratio of the ground state in the central layers depending on its
  polarization and the number of lattice sites.  The changing aspect
  ratio should be observable in time of flight images.  Furthermore,
  we show that the interlayer dipole-dipole interaction reduces the
  excitation energy of local perturbations affecting the development
  of a roton minimum.
\end{abstract}
\maketitle

\section{Introduction}
Layered structures of magnetic materials play a crucial role both in
today's technology and in fundamental physical theories.
Technological examples are aplenty in the magneto-electronic
industries, \eg, hard disks or magnetic sensors.  One theoretical
goal of studying multilayers is to illuminate the elusive theory of
high-$T_c$ superconductivity, where the layered structure appears to
play a crucial role~\cite{LeeNagWen06}.  For a realistic theory of
atomic or molecular multilayers it is, however, vital to include the
dipole-dipole interaction (DDI) between the underlying particles.

The study of magnetic single- and multilayer films has enjoyed a long
history in condensed matter physics (for a recent review, see
Ref.~\cite{DeBMacWhi00} and references therein).  There, an
alternating structure of ferromagnetic and nonmagnetic layers is
deposited on a substrate, \eg, by atomic beam epitaxy.  However,
structural instabilities induced, \eg, by temperature changes and
film thickness variation often complicate experiments in thin films.

Quantum-degenerate dipolar gases have received much attention recently
from both theoretical and experimental studies (for recent reviews,
see Refs.~\cite{LahMenSan09,Bar08}).  Their DDI crucially affects the
ground-state properties~\cite{GorRzaPfa00,YiYou00},
stability~\cite{SanShlZol00,SanShlLew03,Fis06}, and dynamics of the
gas~\cite{YiYou01}.  Furthermore, they offer a route for studying
exciting many-body quantum effects, such as a superfluid-to-crystal
quantum phase transition~\cite{BueDemLuk07},
supersolids~\cite{GorSanLew02} or even topological
order~\cite{MicBreZol06}.  Recent advances in experimental techniques
have paved the way for a Bose-Einstein condensate (BEC) of
\textsuperscript{52}Cr with a magnetic dipole moment $6\mu_B$ (Bohr
magneton $\mu_B$), much larger than conventional alkali
BECs~\cite{GriWerHen05,StuGriKoc05,KocLahMet08}.  Promising candidates
for future dipolar BEC experiments are Er and Dy with even larger
magnetic moments of $7\mu_B$ and $10\mu_B$,
respectively~\cite{BerHanMcC08,LuYouLev10}.  Furthermore, DDI-induced
decoherence and spin textures have been observed in alkali-metal
condensates~\cite{FatRoaDei08,VenLesGuz08}.  Dipolar effects also play
a crucial role in experiments with Rydberg atoms~\cite{VogVitZha06}
and heteronuclear molecules~\cite{NiOspWan10,DeMChoNey11}.  Bosonic
heteronuclear molecules may provide a basis for future experiments on
BECs with dipole moments much larger than in atomic
BECs~\cite{VoiTagCos09}.

\begin{figure}[b]
  \centering
  \includegraphics[width=\linewidth]{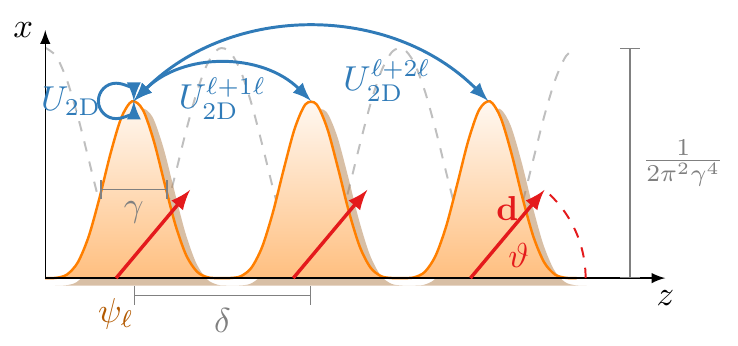}
  \caption{(Color online) Setup of the multilayered dipolar BEC
    polarized along $\bv d$.  An optical lattice along $z$ separates
    the dipolar BEC into 2D layers in the $x$--$y$ plane with distance
    $\delta$.  Apart from the intralayer DDI $U_\text{2D}$, each layer
    interacts with other layers via the interlayer DDI
    $U_\text{2D}^{j\ell}$.}\label{fig:setup}
\end{figure}

In contrast to solid state thin film structures, the layer width and
spacing of BECs in optical lattices are precisely tunable with
external fields.  This makes dipolar BECs a prime candidate for
investigating the effects of DDI in multilayers.  For example, it has
been shown that the DDI stabilizes quasi-two-dimensional ultracold
gases for perpendicular polarization~\cite{Fis06,MueBilHen11} and
enables controlled chemical reactions~\cite{DeMChoNey11}.  Another
intriguing effect is the occurrence of interlayer bound
states~\cite{KlaPikSan10,BarMicRon11,ShiWan09,PotBerWan10,RosBao11,
  VolFedJen11}.  However, it is still unclear to what extend effective
models for multilayers of dipolar BEC at arbitrary polarization are
valid and how interlayer DDI can be detected.

In this article, we investigate the effect of interlayer DDI on the
ground state of the BEC.  We present an effective two-dimensional (2D)
model for an arbitrarily polarized dipolar BEC in a strong
one-dimensional (1D) optical lattice.  Our 2D model offers a clear
advantage for numerical computation of ground state properties
compared to computations for a full three-dimensional (3D)
Gross-Pitaevskii equation (GPE): our computation times reduce to
seconds instead of dozens of hours.  Previously, such
dimension-reduced models have been derived for BECs without
DDI~\cite{BaoTan03,BaoJakMar03,Luca09,NaoMeScWe05,NaoCasMe08,LieSeiYng03,
  YngLieSei04} and with dipolar interactions in a single
layer~\cite{CarMarSpa08,CaiRosLei10}.  We also derive the effective 2D
intra- and interlayer DDI potentials governing the layers of quasi-2D
BECs.  These potentials allow for useful analytical approximations,
which were used in a previous work on multilayer dipolar BECs with
perpendicular polarization~\cite{BarMicRon11}.  We establish that the
2D model is valid by comparing its ground states to ground states of
the 3D GPE for weakly interacting BECs at zero
temperature~\cite{PitStr03}.  We suggest that the interlayer DDI is
observable in the transverse aspect ratio of the central layers after
time of flight expansion.  Moreover, we calculate the Bogoliubov
excitation energies for a transversely homogeneous BEC with contact,
intra- and interlayer DDI.  The interlayer DDI reduces the squared
Bogoliubov energy and, therefore, influences the occurance of a roton
minimum.

In Sec.~\ref{sec:model} we present our 2D model and effective intra-
and interlayer potentials for a dipolar BEC trapped in a strong 1D
optical lattice.  We also present a single mode approximation valid
for the central layers of the BEC.  In Sec.~\ref{sec:validity} we
compare ground states of our model and its single mode approximation
to ground states of the 3D GPE.  We find good agreement between these
ground states, which indicates the validity of our model.  In
Sec.~\ref{sec:aspect} we compute numerically the aspect ratio of the
BEC in the central layer as a function of the number of lattice sites
and polarization direction.  We find a marked change in the aspect
ratio owing to the interlayer DDI, which should be observable in
experiments.  In Sec.~\ref{sec:bogoliubov} we derive the Bogoliubov
dispersion for a transverse homogeneous, multilayered dipolar BEC.  We
conclude in Sec.~\ref{sec:conclusion}.  In App.~\ref{app:derivation}
we give a detailed derivation of the 2D model presented in
Sec.~\ref{sec:model}.

\section{Effective 2D model}\label{sec:model}
We consider a dilute dipolar BEC at zero temperature trapped in a
transverse harmonic potential $V_\text{ho}(x, y) =
\tfrac{m\omega^2}{2} (x^2 + y^2)$ and a longitudinal optical lattice
$V_\text{o}(z) = V_0 \sin^2(k_lz)$.  Here, $m$ is the particle mass,
$\omega$ the trap frequency, $V_0$ the lattice height, and $k_l$ the
wave number of the lattice laser.  We focus on atomic BECs with a
magnetic dipole moment but it is straightforward to extend the
analysis to degenerate bosonic gases with electric dipole moments.  We
assume that an external field polarizes the atoms along a normalized
axis $\bv d = (d_x, d_y, d_z) = (\cos\phi\sin\vartheta,
\sin\phi\sin\vartheta, \cos\vartheta)$ with $\phi$ and $\vartheta$ the
azimuthal and polar angles, respectively.  Then the dipole-dipole
interaction (DDI) is described by
\begin{equation}\label{eq:U_dd}
  U_\text{dd}(\bv r) = \frac{c_\text{dd}}{4\pi} \frac{|\bv r|^2 -
    3(\bv d\cdot \bv r)^2}{|\bv r|^5},
\end{equation}
where $c_\text{dd} = \mu_0 D^2$ with $\mu_0$ is the magnetic vacuum
permeability and $D$ the dipole moment (for electric dipoles $c_{dd} =
D^2/\epsilon_0$, where $\epsilon_0$ is the vacuum permittivity).  We
note that it is possible to modify the DDI strength $c_\text{dd}$ by
means of a rotating magnetic field~\cite{GioGoePfa02}.

At zero temperature, a weakly interacting BEC is described by the
GPE~\cite{PitStr03}.  For simplicity, we introduce dimensionless
quantities by rescaling lengths with the lattice distance $\delta =
\pi/k_l$, that is, $\bv r \rightarrow \bv r \delta$, energies with
$\hbar^2/m\delta^2 = 2E_r/\pi^2$ ($E_r$ is the recoil energy), and the
wave function of the gas with the central density $n(0)$, $\psi
\rightarrow \psi \sqrt{n(0)}$.  In these units the normalization of
the wave function is $\int d^3\bv r |\psi(\bv r, t)|^2 =
N/n(0)\delta^3$ with $N$ the total number of atoms.  Away from shape
resonances, the wave function $\psi = \psi(\bv r, t)$ of the dipolar
BEC is governed by the GPE~\cite{MarYou98,YiYou00,DebYou01}
\begin{equation}\label{eq:GPE}
  i\partial_t \psi = \left[-\frac{1}{2} \nabla^2 + V_\text{ho} +
    V_\text{o} + (g - g_d) |\psi|^2 + V_\text{dd}
  \right] \psi.
\end{equation}
Here, $g = 4\pi a_s n(0) \delta^2$ is the dimensionless contact
interaction strength with $a_s$ the s-wave scattering length and $g_d
= mc_\text{dd}n(0) \delta^2/3\hbar^2$ is the dimensionless DDI
strength.  Furthermore, $V_\text{ho}(\bv\rho) =
(m^2\omega^2\delta^4/2\hbar^2) \bv\rho^2$ with $\bv\rho = (x,y)$ and
$V_\text{o}(z) = (\bar V_0 \pi^2/2) \sin^2(\pi z)$, where $\bar V_0$
is the lattice amplitude in units of the recoil energy $E_r$.  The
nonlocal dipolar potential $V_\text{dd}$ is given by
\begin{equation}\label{eq:V_dd}
  V_\text{dd}(\bv r) = -3g_d \partial_{\bv d\bv d}
  \int d^3\bv r' U_\text{3D}(\bv r - \bv r') |\psi(\bv r', t)|^2
\end{equation}
with the kernel $U_\text{3D}(\bv r) = 1/4\pi|\bv r|$ and the notation
$\partial_{\bv d} = \bv d\cdot \bv\nabla$, $\partial_{\bv d\bv d}
= \partial_{\bv d}^2$.

\subsection{Coupled modes}
For strong optical lattices we derive an effective 2D equation for the
wave function on each lattice site.  This is possible because a strong
optical lattice with $V_0 \gg \hbar\omega$ causes the BEC to form
layers separated by the lattice distance $\delta$
(cf. Fig.~\ref{fig:setup})~\cite{BurCatFor02,PitStr03}.  We assume
that the axial extend $\gamma$ of the BEC in each layer is much larger
than the s-wave scattering length.  Additionally, in the quasi-2D
regime $\gamma^{-2} \gg |g-g_d|$~\cite{PetHolShl00}.  This condition
allows us to approximate the optical lattice as a train of harmonic
potentials and the axial wave function as its ground state.  Then the
wave function separates into $\psi(\bv r, t) = \eu^{-\im t/2\gamma^2}
\sum_\ell \psi_\ell(\bv\rho, t)
w_\ell(z)$~\cite{BaoJakMar03,PitStr03,CaiRosLei10}. The sum extends
over all lattice sites $\ell$.  Under our assumptions the axial wave
function on each site $\ell$ at position $z_\ell$ is described by a
Gaussian $w_\ell(z) = w(z-z_\ell) = (1/\pi\gamma^2)^{1/4}
\eu^{-(z-z_\ell)^2/2\gamma^2}$; the Gaussians do not mutually overlap
($\int dz w_\ell(z) w_j(z) \simeq 0$ for $\ell\neq j$).  In the
quasi-2D limit $\gamma^{-2} = \sqrt{\bar V_0} \pi^2$.  More generally,
in a homogeneous BEC it is also possible to treat the layer width
$\gamma$ as a variational parameter that minimizes the
Gross-Pitaevskii energy functional~\cite{Fis06}.  By inserting this
wave function into Eq.~\eqref{eq:GPE} and integrating out the $z$
direction we obtain the following equation for the radial wave
function $\psi_\ell=\psi_\ell(\bv\rho, t)$ at site $\ell$
\begin{equation}\label{eq:GPE-2d}
  \im \partial_t \psi_\ell = \left[-\frac{1}{2} \nabla^2 + V_\text{ho} +
    \bigl[\bar g - \bar g_d \bigl(1 - 3d_z^2\bigr)
    \bigr] |\psi_\ell|^2 + V_\text{2D}^\ell \right] \psi_\ell.
\end{equation}
Here, $\bar g = g/\sqrt{2\pi}\gamma$ and $\bar g_d =
g_d/\sqrt{2\pi}\gamma$ are the effective 2D interactions strengths.
In the remainder of this article we neglect strongly suppressed terms
in the effective DDI potential $V_\text{2D}^\ell$ (see
Appendix~\ref{app:derivation} for details).  We find the following
expression for its Fourier transform $\hat V_\text{2D}^\ell(\bv k) =
\mathcal{F}[V_\text{2D}^\ell](\bv k)$ with $\bv k = k(\cos\varphi,
\sin\varphi)$
\begin{equation}\label{eq:V_2d}
  \begin{split}
    \hat V_\text{2D}^\ell(\bv k) &= 3g_d \sum_{j} \Bigl(\left[(d_x
      \cos\varphi + d_y \sin\varphi)^2 - d_z^2
    \right] \hat U_\text{even}^{j\ell}(k)\\
    &\quad + 2\im d_z (d_x\cos\varphi + d_y\sin\varphi) \hat
    U_\text{odd}^{j\ell}(k) \Bigr) \widehat{|\psi_j|^2}(\bv k).
  \end{split}
\end{equation}
Here,
\begin{align}
  \hat U_\text{even}^{j\ell}(k) &= \frac{k}{4}
  \eu^{-\tfrac{\delta_{\ell j}^2}{2\gamma^2}} \left[
    \eta\left(\frac{\gamma^2 k + \delta_{\ell
          j}}{\sqrt 2\gamma}\right) + \eta\left(\frac{\gamma^2 k -
        \delta_{\ell j}}{\sqrt 2\gamma}\right) \right],\label{eq:U_even}\\
  \hat U_\text{odd}^{j\ell}(k) &= \frac{k}{4}
  \eu^{-\tfrac{\delta_{\ell j}^2}{2\gamma^2}} \left[
    \eta\left(\frac{\gamma^2 k + \delta_{\ell
          j}}{\sqrt 2\gamma}\right) -\eta\left(\frac{\gamma^2 k -
        \delta_{\ell j}}{\sqrt 2\gamma}\right)
  \right],\label{eq:U_odd}
\end{align}
where $\delta_{\ell j} = (\ell-j)$, $\eta(x) = \exp(x^2) \erfc(x)$ and
$\erfc(x) = 1 - \erf(x)$ is the complementary error function.

\begin{figure}
  \includegraphics[width=\linewidth]{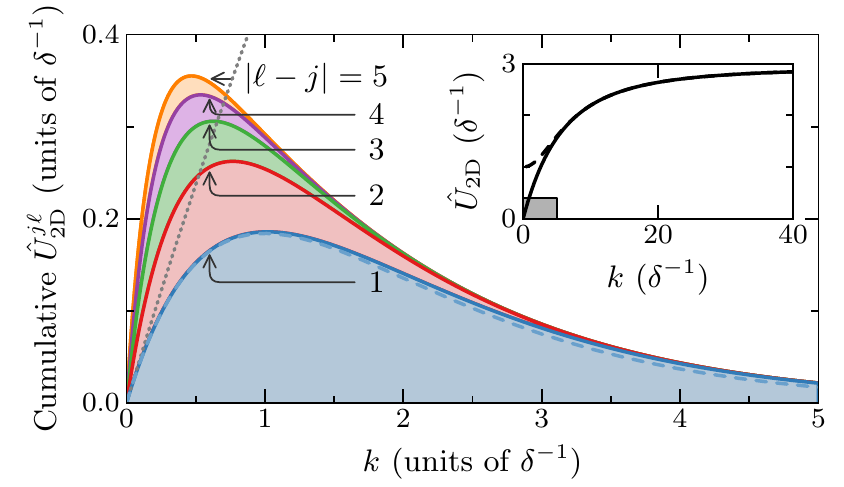}
  \caption{(Color online) Cumulative interlayer DDI $\hat
    U_\text{2D}^{j\ell}$ for \textsuperscript{52}Cr at different layer
    separations $|\ell-j|$.  The solid lines show the interlayer DDI
    [Eq.~\eqref{eq:U_2d}], whereas the dashed line shows the
    approximation Eq.~\eqref{eq:U_2d-approx} for nearest neighbors
    [Eq.~\eqref{eq:U_2d-approx} is indistinguishable from the solid
    lines for larger distances].  The dotted line indicates the
    intralayer DDI.  The inset shows the intralayer DDI and the
    rectangle within indicates the extend of the main panel.  We set
    $V_0 = 30E_r$.}\label{fig:potentials}
\end{figure}

The effective dipolar interaction $\hat V_\text{2D}^{\ell}$
[Eq.~\eqref{eq:V_2d}] contains both an intralayer DDI and an
interlayer DDI.  The intralayer DDI are the terms in
Eq.~\eqref{eq:V_2d} with $\ell=j$.  By setting $\ell=j$ in
Eqs.~\eqref{eq:U_even}-\eqref{eq:U_odd} we find that each layer
experiences the effective DDI potential of a quasi-2D dipolar
BEC~\cite{CaiRosLei10}.  The interlayer DDI are the terms in
Eq.~\eqref{eq:V_2d} with $\ell\neq j$.  For perpendicular polarization
($d_z=1$, $d_x=d_y=0$) we recover the interlayer DDI potential
discussed, \eg, in Ref.~\cite{BarMicRon11}.  If the layer distance is
much larger than the layer width ($|\delta_{\ell j}| \gg \gamma$),
$\eta(x\rightarrow+\infty)$ vanishes and the moduli of the kernels
$|\hat U_\text{even}^{j\ell}|$ and $|\hat U_\text{odd}^{j\ell}|$
become identical.  Because of our assumption that $\delta \gg \gamma$,
this is fulfilled for the interlayer DDI between any two distinct
sites.  As a consequence, we split the total effective DDI potential
into a sum of intralayer and interlayer terms
\begin{equation}\label{eq:V_2d-split}
  \begin{split}
    \hat V_\text{2D}^\ell(\bv k) &= 3g_d [(d_x\cos\varphi +
    d_y\sin\varphi)^2 - d_z^2] \hat
    U_\text{2D}(k) \widehat{|\psi_\ell|^2}(\bv k)\\
    &\quad + 3g_d \sum_{j\neq\ell} [d_x\cos\varphi + d_y\sin\varphi - \im
    d_z \sgn(\delta_{\ell j})]^2\\
    &\quad\times \hat U_\text{2D}^{j\ell}(k) \widehat{|\psi_j|^2}(\bv k),
  \end{split}
\end{equation}
where $\sgn(x)$ is the sign of $x$.  The kernels of this potential are
$\hat U_\text{2D} = 2\hat U_\text{2D}^{00}$ and
\begin{equation}\label{eq:U_2d}
  \hat U_\text{2D}^{j\ell}(k) = \frac{k}{4} \eu^{-\frac{\delta_{\ell
        j}^2}{2\gamma^2}} \eta\left(\frac{\gamma^2 k - |\delta_{\ell
        j}|}{\sqrt 2\gamma} \right).
\end{equation}
In the limit of negligible layer width ($\gamma \ll |\delta_{\ell
  j}|$) the interlayer DDI in Eq.~\eqref{eq:U_2d} can be approximated
by
\begin{equation}\label{eq:U_2d-approx}
  \hat U_\text{2D}^{j\ell}(k) \simeq \frac{k}{2}
  \eu^{-|\delta_{\ell j}| k}\quad (\ell\neq j).
\end{equation}
This approximation becomes an identity in the limit $\gamma
\rightarrow 0$ and nonzero $|\delta_{\ell j}|$.  The second line of
Eq.~\eqref{eq:V_2d-split} is the interlayer DDI potential for
arbitrary polarization direction.  Inserting
approximation~\eqref{eq:U_2d-approx} into Eq.~\eqref{eq:V_2d-split}
for perpendicular polarization, we recover the interlayer DDI
potential used in Refs.~\cite{PotBerWan10,BarMicRon11}.  We expect our
generalized interlayer DDI potential to be valid for bosons as well as
fermions because fermions in different layers occupy different quantum
states.

The kernel of the interlayer DDI potential $\hat
U_\text{2D}^{j\ell}(k)$ is shown in Fig.~\ref{fig:potentials} as a
cumulative plot over the five nearest lattice sites.  For comparison
we also show the intralayer DDI.  Although not shown in
Fig.~\ref{fig:potentials}, we established that for realistic
parameters the potentials $\hat U_\text{even}^{j\ell}$ and $\hat
U_\text{odd}^{j\ell}$ (for $\ell\neq j$) are indistinguishable from
$\hat U_\text{2D}^{j\ell}$ at the plot resolution.  For interlayer
interactions beyond nearest neighbors the approximation for $\hat
U_\text{2D}^{j\ell}$ in Eq.~\eqref{eq:U_2d-approx} becomes
indistinguishable from Eq.~\eqref{eq:U_2d}.  The interlayer DDI is
linear in momentum for long wavelengths and drops exponentially for
short wavelengths.  It has been shown that this behavior leads to very
weakly bound states in bilayer
systems~\cite{VolFedJen11,VolZinFed11,BarMicRon11,RosBao11}.
According to Eq.~\eqref{eq:V_2d-split} its sign is determined by the
polarization direction.  The interlayer and intralayer DDI for
predominantly perpendicular polarization ($\vartheta < \pi/4$) is
attractive in momentum space for all $\bv k$, whereas the interlayer
DDI for predominantly parallel polarization ($\vartheta > \pi/4$)
becomes repulsive for some $\bv k$ around the major axis with $\varphi
= \phi$.

\subsection{Single mode approximation}
If we assume that the the BEC densities in each layer vary little over
the central sites, we can simplify the 2D model to a single equation
for the central site wave function $\psi_0(\bv\rho)$.  This assumption
is reasonable for large lattices and we will test its validity in
Sec.~\ref{sec:validity}.  The single wave function $\psi_0(\bv\rho)$
approximates the wave functions in all lattice sites far from the
boundaries.  Consequently, we replace the effective dipolar potential
$\hat V_\text{2D}^\ell(\bv k)$ [Eq.~\eqref{eq:V_2d-split}] by the
site-local potential
\begin{equation}\label{eq:V_2d-single}
  \begin{split}
    \hat V_\text{2D}(\bv k) &= 3g_d \Bigl( [(d_x\cos\varphi +
    d_y\sin\varphi)^2 - d_z^2] \hat
    U_\text{2D}(k)\\
    &\quad + \sum_{j\neq 0} [d_x\cos\varphi + d_y\sin\varphi - \im
    d_z \sgn(j)]^2 \hat U_\text{2D}^{j0}(k) \Bigr)\\
    &\quad\times \widehat{|\psi_0|^2}(\bv k).
  \end{split}
\end{equation}
Inserting the inverse Fourier transform of Eq.~\eqref{eq:V_2d-single}
into Eq.~\eqref{eq:GPE-2d} we are left with the uncoupled equation
\begin{equation}\label{eq:GPE-2d-sma}
  \im \partial_t \psi_0 = \left[-\frac{1}{2} \nabla^2 + V_\text{ho} +
    \bigl[\bar g - \bar g_d \bigl(1 - 3d_z^2\bigr)
    \bigr] |\psi_0|^2 + V_\text{2D} \right] \psi_0
\end{equation}
for the central site wave function $\psi_0 = \psi_0(\bv\rho)$.  We
assume a lattice that is symmetric around the central site so that the
dipole terms linear in $d_z$ in Eq.~\eqref{eq:V_2d-single} vanish
after summation.  Using Eq.~\eqref{eq:U_2d-approx} for $\hat
U_\text{2D}^{j0}$ we can perform the summation in
Eq.~\eqref{eq:V_2d-single} and find
\begin{equation}\label{eq:V_2d-summed}
  \begin{split}
    \hat V_\text{2D}(\bv k) &\simeq 3g_d [(d_x\cos\varphi +
    d_y\sin\varphi)^2 - d_z^2]\\
    &\quad \times \bigl[\hat U_\text{2D}(k) + \hat
    U_\text{2D}^{N_s^*}(k) \bigr] \widehat{|\psi_0|^2}(\bv k)
  \end{split}
\end{equation}
with
\begin{equation}
  \hat U_\text{2D}^{N_s^*}(k) = k \left(\frac{1 - \eu^{-(N_s^* + 1)k/2}}{1 -
      \eu^{-k}} - 1 \right).
\end{equation}
Here, we summed over $N_s^*$ central lattice sites.  In the limit of
an infinite lattice the maximum of $U_\text{2D}^{N_s^*}$ moves towards
$k=0$ with $\lim_{k\rightarrow 0} \hat U_\text{2D}^{\infty}(k) = 1$.
Therefore, the total DDI potential for an infinite stack of BECs does
not vanish anymore at $k = 0$ (dashed line in the inset of
Fig.~\ref{fig:potentials}).  However, this is a pathological case
because for any finite $N_s$ the total DDI potential vanishes at $k=0$
and our assumption of slowly varying wave functions breaks down
towards the boundary.

\section{Validity of the 2D model}\label{sec:validity}
In this section, we investigate the validity of the effective 2D model
for multilayered dipolar BECs introduced in Sec.~\ref{sec:model}.  To
this end we computed ground states for the 3D GPE [Eq.~\eqref{eq:GPE}]
~\cite{BaoCaiWan10}, the coupled 2D model
[Eq.~\eqref{eq:GPE-2d}]~\cite{Bao04}, and the single mode 2D model
[Eq.~\eqref{eq:GPE-2d-sma}]~\cite{BaoDu04} using the normalized
gradient flow (imaginary time) method.  For the time discretization we
used backward Euler finite difference~\cite{BaoDu04}.  For the spatial
discretization we employed the sine pseudospectral~\cite{BaoCaiWan10}
and the Fourier pseudospectral methods~\cite{CaiRosLei10} for the 3D
GPE and the 2D models, respectively. For the 3D computation we assumed
that the wave function vanishes at the boundaries.  We integrated the
3D ground states over the individual lattice sites to find the $N_s$
densities $|\psi_\ell^\text{3D}(\bv\rho)|^2 =
\int_{\delta(\ell-1/2)}^{\delta(\ell+1/2)} dz |\psi(\bv r)|^2$.  To
determine the validity of the 2D model we compared the 2D ground states
$\psi_\ell(\bv\rho)$ to $\psi_\ell^\text{3D}(\bv\rho)$.  Using the
single mode approximation reduced the computation times drastically:
typically to less than a minute, compared to $2$--$3$ hours for the
coupled equations and $\sim 1$ day for the 3D GPE.  In this section we
only consider polarization in the $x$--$z$ plane, that is $\bv d =
(\sin\vartheta,0,\cos\vartheta)$ (cf. Fig.~\ref{fig:setup}).  Because
the external potential is radially symmetric, this simplification
corresponds to choosing the transverse projection of the polarization
direction as the $x$ axis.

\begin{figure}
  \centering\includegraphics[width=\linewidth]{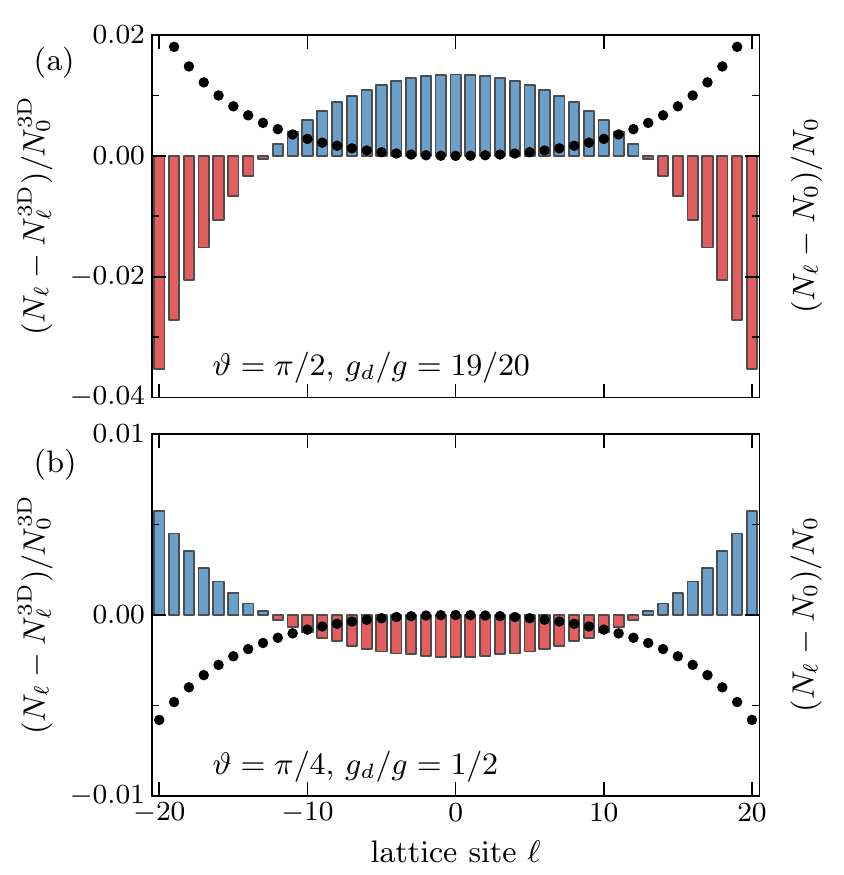}
  \caption{(Color online) Relative particle number difference between
    GPE ground state and the 2D model [Eq.~\eqref{eq:GPE-2d}] for
    individual lattice sites.  The particle numbers are relative to
    the particle number in the central layer $N_0^\text{3D}$ (bars).
    The discs indicate the particle number difference in the 2D model
    relative to the central site (right axis label).  The parameters
    are $N_s = 61$ lattice sites with $V_0 = 20 E_r$, $E_r/\hbar\omega
    = 60$, and $g = 100
    \sqrt{2E_r/\hbar\omega\pi^2}$.}\label{fig:proj1d}
\end{figure}

To compare the axial profiles of the coupled 2D and 3D ground states
we computed the relative particle numbers in each lattice site.
Because of the long range of the DDI, we observe fairly pronounced
boundary effects in the 3D computations for strong dipolar
interactions $g_d \simeq g$.  For this reason we omit the $N_b$
outermost lattice sites in the overall normalization.  Then the relative
number of particles in site $\ell$ for the 2D model is given by
$N_\ell = \int d^2\bv\rho |\psi_\ell(\bv\rho)|^2 /
\sum_{j=-N_s+N_b}^{N_s-N_b} \int d^2\bv\rho |\psi_\ell(\bv\rho)|^2$
(the relative particle number $N_\ell^\text{3D}$ for the 3D GPE
follows by replacing $|\psi_\ell|^2$ with $|\psi_\ell^\text{3D}|^2$).
Figure~\ref{fig:proj1d} shows the particle number difference
$(N_\ell^\text{3D}-N_\ell) / N_0^\text{3D}$ relative to the particle
number at the central lattice site.  Although the number difference
varies slightly over the central lattice sites, the difference between
the GPE and the 2D model Eq.~\eqref{eq:GPE-2d} remains smaller than
$4\%$ and $1\%$ for the two parameter sets, respectively.

\begin{figure}
  \centering\includegraphics[width=\linewidth]{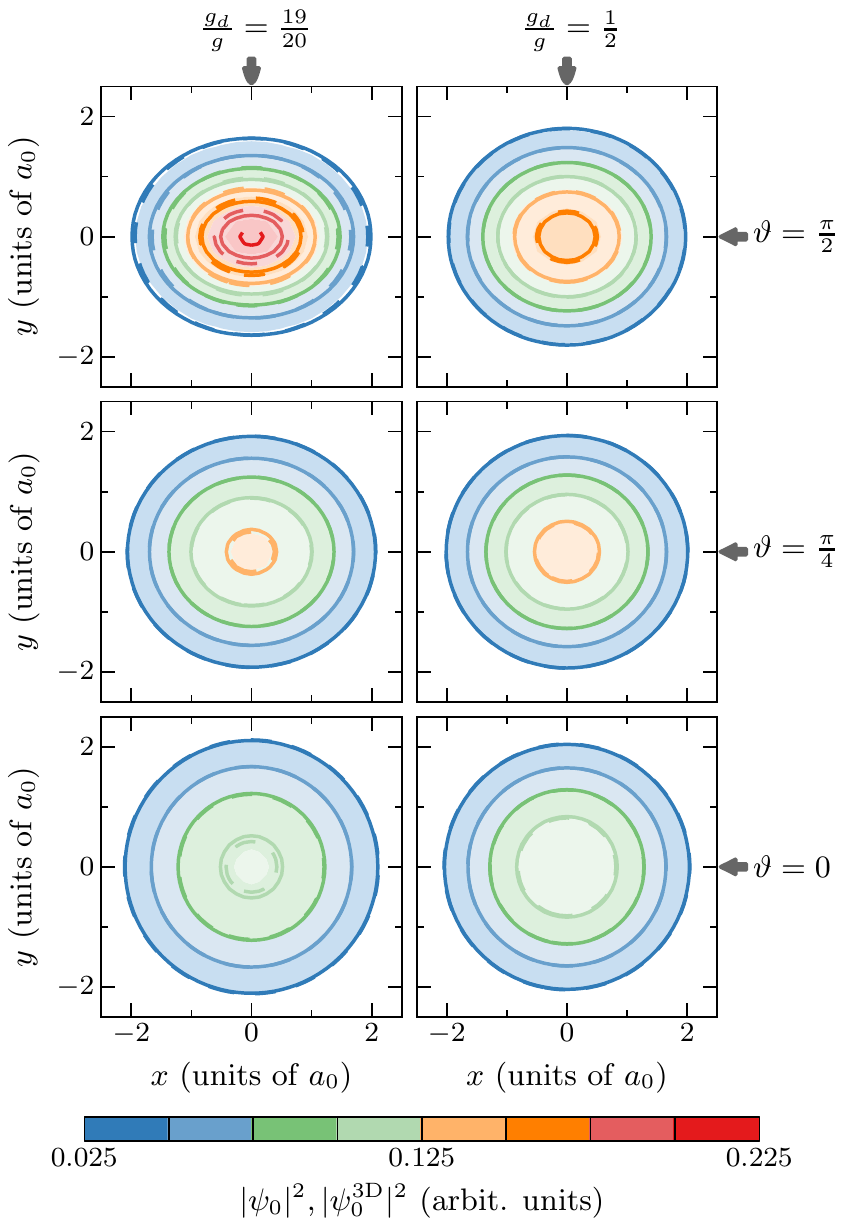}
  \caption{(Color online) Ground state densities of the central
    lattice site for various DDI strengths and polarization angles.
    The filled surfaces are the projection of the central site of the
    GPE results, whereas the solid (dashed) contour lines are the
    ground states of the coupled (single mode) 2D
    equation~\eqref{eq:GPE-2d}.  The plotted densities are all
    normalized to $1$.  The coupled and single mode results are almost
    indistinguishable except in the top left panel.  The parameters
    are as in Fig.~\ref{fig:proj1d}.  The plots use the
    magnetic length $a_0 = \sqrt{\hbar/m\omega}$ as length
    unit.}\label{fig:proj2d}
\end{figure}

Next we compared the density profiles of the central lattice site
$|\psi_0(\bv\rho)|^2$ for the coupled and single mode models with
$|\psi_0^\text{3D}(\bv\rho)|^2$.  The sum of the densities of the
coupled 2D and the total density of the 3D GPE are normalized to a
function proportional of the total particle number $\mathcal{N}(N)$.
However, in the single mode approximation we only consider a single
wave function which has, consequently, a normalization less than
$\mathcal{N}$.  If the BEC density were the same in all layers, the
normalization of this single wave function would be
$\mathcal{N}/\sqrt{N_s}$.  Because the density varies slightly across
layers, instead we chose to normalize the single mode density to the
particle number in the central layer of the GPE.  The ground state
densities for various DDI strengths and polarization angles are shown
in Fig.~\ref{fig:proj2d}.  We find that both the coupled and single
mode models describe the ground state well for any polarization.  We
only observe a slight difference between the models for strong DDI on
the order of the contact interaction and parallel polarization (top
left panel in Fig.~\ref{fig:proj2d}).  This means that even the single
mode approximation describes the ground state of the multilayer
dipolar BEC well.  Its accuracy diminishes for strong DDI because the
true densities vary sufficiently strongly over the central lattice
sites.

\section{Interlayer-DDI-induced change of the aspect
  ratio}\label{sec:aspect}
The interlayer DDI can cause observable effects in multilayered
dipolar BECs.  This becomes apparent from Fig.~\ref{fig:potentials}.
The strength of the interlayer DDI is comparable to the strength of
the intralayer DDI at wavelengths larger than $\delta$.  We expect
that the anisotropy of the DDI for $\vartheta > 0$ leads to a change
in the aspect ratio of a quasi-2D dipolar BEC in the central layer of
a stack of dipolar BECs.  In this section, we investigate these
effects numerically using the single mode approximation for the
central layer.

\begin{figure}
  \centering\includegraphics[width=\linewidth]{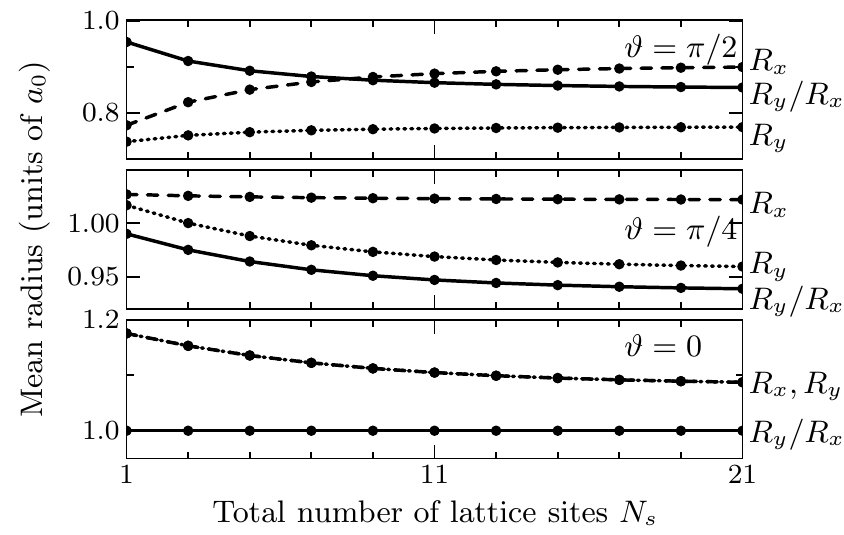}
  \caption{(Color online) Mean radii and aspect ratio of the central
    BEC layer as a function of the number of lattice sites.  The
    different panels correspond to different polarization angles.  The
    interlayer DDI has a noticeable effect over several lattice sites.
    The lines are marked at the right and are only to guide the eye.
    The parameters are as in Fig.~\ref{fig:proj1d} with $g_d/g =
    19/20$.}\label{fig:interlayer}
\end{figure}

To determine the mean radii of the central layer first we computed
ground state densities for a varying number of lattice sites at a
constant normalization.  We calculated the mean radii as
\begin{equation}
  R_{\alpha}^2 = \int d^2\bv\rho \alpha^2 |\psi_0(\bv\rho)|^2,
  \quad(\alpha = x, y).
\end{equation}
The aspect ratio of the central layer is then given by $R_y/R_x$.
Magnetostriction causes the dipolar BEC to expand along the
polarization direction~\cite{LahMenSan09,CaiRosLei10}.
Figure~\ref{fig:interlayer} shows the aspect ratio as well as the
individual mean radii of the BEC as a function of the number of
lattice sites $N_s$.  The case $N_s=1$ corresponds to a single layer
dipolar BEC.  We observe that the interlayer DDI causes an additional
reduction in the aspect ratio depending on the number of lattice sites
and polarization angle.  For perpendicular polarization the aspect
ratio remains unchanged because the DDI is isotropic.  However, the
individual radii decrease.  We have also computed aspect ratios for a
stronger lattice with $V_0 = 40E_r$ and observed a similar dependence
of the mean radii on $N_s$.  For this stronger lattice and
$\vartheta=\pi/4$ the aspect ratio was closer to $1$ and its change
slightly smaller than at $V_0=20E_r$.  For perpendicular polarization
the mean radii and aspect ratio were nearly indistinguishable from the
top panel in Fig.~\ref{fig:interlayer}.  The DDI-induced change of
aspect ratio has been observed in a single layer
\textsuperscript{52}Cr via time of flight
expansion\cite{StuGriKoc05,LahKocFro07}.  We suggest that the
dependence of the aspect ratio on $N_s$ could also be observed via
time of flight expansion.  To observe the central layers, in this
experiment the outer layers would have to be removed on a time scale
short enough to suppress equilibration, \eg, with additional lasers
focused on the outer layers.  This is followed immediately by time of
flight expansion of the BEC.  The observable effect is largest for
parallel polarization $\vartheta = \pi/2$.

\section{Bogoliubov excitations}\label{sec:bogoliubov}
In this section we investigate the influence of interlayer DDI on the
excitation spectrum of a layered quasi-2D dipolar BEC.  In particular,
we consider local density fluctuations of the layered BEC and derive
their Bogoliubov energy.  Their Bogoliubov energy can assume imaginary
values for suitable parameters, which indicates the onset of a
dynamical instability that leads to exponential growth of excitations.

To determine the Bogoliubov energy we consider small perturbations
around the ground state of Eq.~\eqref{eq:GPE-2d}.  For simplicity we
assume a vanishing transverse harmonic potential $V_\text{ho} = 0$ and
homogeneous density $\nu$ in each layer.  For an optical lattice with
$N_s$ sites $\nu = 1/N_s$.  A stationary state of the effective 2D
GPE~\eqref{eq:GPE-2d} is given by $\psi_\ell(\bv\rho, t) =
\psi_\ell(t) = \eu^{-\im\mu t} \sqrt\nu$ with the chemical potential
\begin{equation}
  \mu = [\bar g - \bar g_d (1 - 3d_z^2)] \nu.
\end{equation}
Now we add a local perturbation $\xi_\ell(\bv\rho, t)$ to the
stationary state $\psi_\ell(t)$, that is, $\psi_\ell(\bv\rho, t) =
\eu^{-\im\mu t} [\sqrt\nu + \xi(\bv\rho, t)]$.  We expand the
perturbation in a plane wave basis as $\xi_\ell(\bv\rho, t) = (1/2\pi)
\int d^2\bv q \bigl( u_{\bv q\ell}\eu^{\im (\bv q\cdot\bv\rho -
  \omega_{\bv q} t)} + v_{\bv q\ell}^* \eu^{-\im (\bv q\cdot\bv\rho -
  \omega_{\bv q} t)} \bigr)$ and insert $\psi_\ell(\bv\rho, t)$ into
Eq.~\eqref{eq:GPE-2d}.  Here, $\omega_q$ are the excitation
frequencies of quasimomentum $\bv q$ and $u_{\bv q\ell}$, $v_{\bv
  q\ell}$ are the mode functions in layer $\ell$.  Keeping terms
linear in the excitations $u_{\bv q\ell}$ and $v_{\bv q\ell}$ we find
the Bogoliubov-de Gennes equations for perpendicular polarization
\begin{align}
  \begin{split}
    \omega_{\bv q} u_{\bv q\ell} &= \frac{q^2}{2} u_{\bv q\ell} +
    \nu (\bar g + 2\bar g_d) (u_{\bv q\ell} + v_{\bv q\ell})\\
    &\quad - g_d\nu \sum_j \hat U_\text{2D}^{j\ell}(q) (u_{\bv qj} +
    v_{\bv qj}),
  \end{split}\label{eq:u_q}\\
  \begin{split}
    - \omega_{\bv q} v_{\bv q\ell} &= \frac{q^2}{2} v_{\bv q\ell} +
    \nu (\bar g + 2\bar g_d) (v_{\bv q\ell} + u_{\bv q\ell})\\
    &\quad - g_d\nu\sum_j \hat U_\text{2D}^{j\ell}(q) (v_{\bv qj} +
    u_{\bv qj}).
  \end{split}\label{eq:v_q}
\end{align}
Excitations in layer $\ell$ are coupled to excitations in all layers
through the interlayer DDI.  However, the interlayer DDI drops
exponentially with the distance [cf. Fig.~\ref{fig:potentials} and
Eq.~\eqref{eq:U_2d}].  Therefore, first we only take into account
nearest neighbor interactions $|\ell-j| \leq 1$.  Then the matrix of
the system of Eqs.~\eqref{eq:u_q}--\eqref{eq:v_q} becomes tridiagonal
and can be solved for its eigenenergies.  The resulting Bogoliubov
energy $E_B(\bv q) = \omega_{\bv q}$ is determined by
\begin{equation}\label{eq:E_B-perp}
  \begin{split}
    E_B^2(q) &= \frac{q^2}{2} \biggl[\frac{q^2}{2} + 2(\bar g
    + 2\bar g_d)\nu\\
    &\quad - 3 g_d\nu \hat U_\text{2D}(q)
    - 12 g_d \nu \hat U_\text{2D}^{\ell+1,\ell}(q) \biggr].
  \end{split}
\end{equation}
Because $\hat U_\text{2D}^{j\ell}(q)$ vanishes for zero quasimomentum,
the speed of sound $c = \lim_{q\rightarrow 0} \partial E_B(q)/\partial
q = \sqrt{\bar g\nu + 2\bar g_d\nu}$ is not influenced by the
interlayer DDI.  Only the intralayer DDI increases the speed of sound
via its zero momentum mode.

Now we generalize the Bogoliubov energy in multilayer dipolar BECs to
arbitrary polarization.  After inserting the expansion of the 2D wave
functions into Eq.~\eqref{eq:GPE-2d} we find the squared Bogoliubov
energy
\begin{equation}\label{eq:E_B}
  \begin{split}
    E_B^2(\bv q) &= \frac{q^2}{2} \biggl[\frac{q^2}{2} + 2 [\bar g
    - \bar g_d(1 - 3d_z^2)]\nu\\
    &\quad + 6 g_d\nu \hat W_\text{2D}^{\ell\ell}(\bv q)
    - 12 g_d \nu \bigl|\hat W_\text{2D}^{\ell+1,\ell}(\bv
    q)\bigr| \biggr].
  \end{split}
\end{equation}
Here, $\hat W_\text{2D}^{j\ell}(\bv q) = [(d_x\cos\varphi +
d_y\sin\varphi)^2 - d_z^2] \hat U_\text{2D}^{j\ell}(q)$ in polar
coordinates $\bv q = q(\cos\varphi, \sin\varphi)$.  In general, this
excitation energy is anisotropic but mirror symmetric around the
polarization direction projected onto the $x$--$y$ plane.
Interestingly, the interlayer interaction always reduces the
Bogoliubov energy compared to the Bogoliubov energy of a dipolar BEC
with only intralayer DDI.  This means that interlayer DDI drives the
BEC closer towards an instability regardless of the polarization
direction.

We gain qualitative insight into instabilities by looking at the
dipole-dominated regime with $g/g_d \rightarrow 0$.  Setting $\bar g =
0$ in Eq.~\eqref{eq:E_B} we see that the contact interaction terms
becomes attractive for polarization angles $d_z^2 = \cos^2\vartheta <
1/3$.  Because the DDI terms (last line) in Eq.~\eqref{eq:E_B} vanish
at $\bv q = 0$, this leads to imaginary Bogoliubov energies at low
quasimomenta $\bv q$.  The dipole-dominated quasi-2D BEC ground state
is not stable in this regime.  However, a repulsive s-wave interaction
$g > g_d(1-3d_z^2)$ prevents this type of instability.  For repulsive
contact interaction ($\cos^2\vartheta > 1/3$) another instability of
the dipole-dominated quasi-2D BEC occurs at nonzero quasimomenta.  For
perpendicular polarization a sufficiently large negative intralayer
DDI term in Eq.~\eqref{eq:E_B-perp} (large $g_d\nu$) compensates the
positive free energy and local terms (first line).  This leads to an
instability in the cross-over regime from quasi-2D to 3D~\cite{Fis06}.
The interlayer DDI term in Eq.~\eqref{eq:E_B-perp} shifts the
instability region to smaller quasimomenta.  Because in the present
article we only consider quasi-2D BECs, we refer to an upcoming
article investigating instabilities in the 2D--3D cross-over
regime~\cite{RosFis11}.

\begin{figure}
  \centering\includegraphics[width=\linewidth]{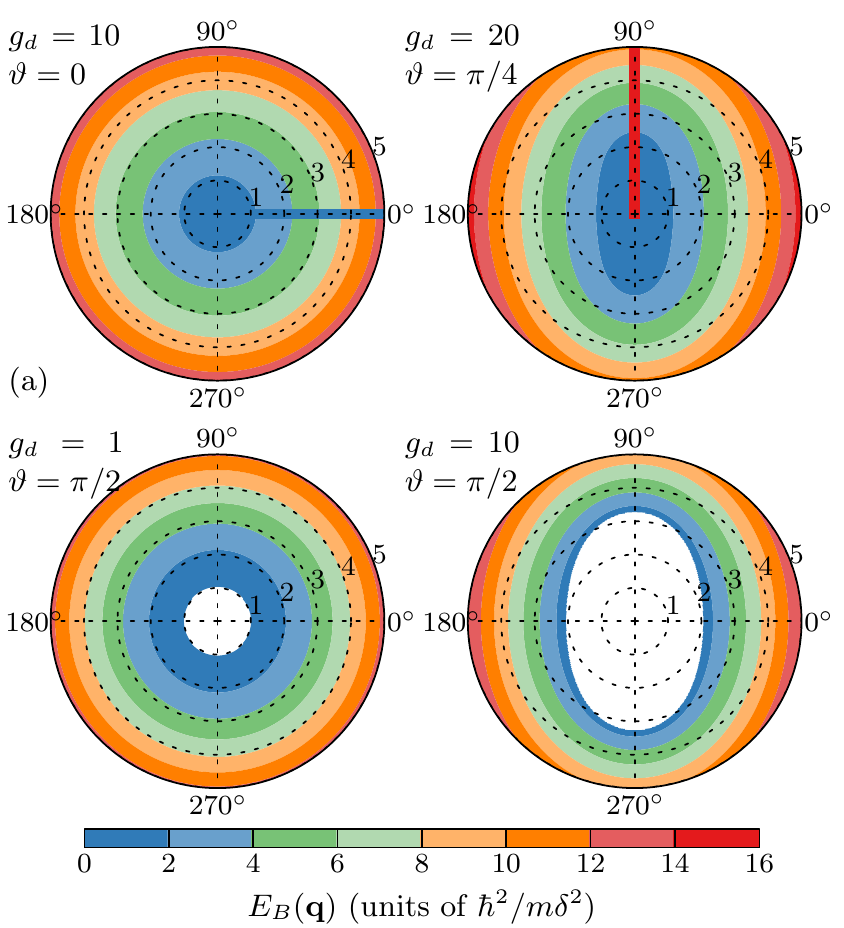}
  \includegraphics[width=\linewidth]{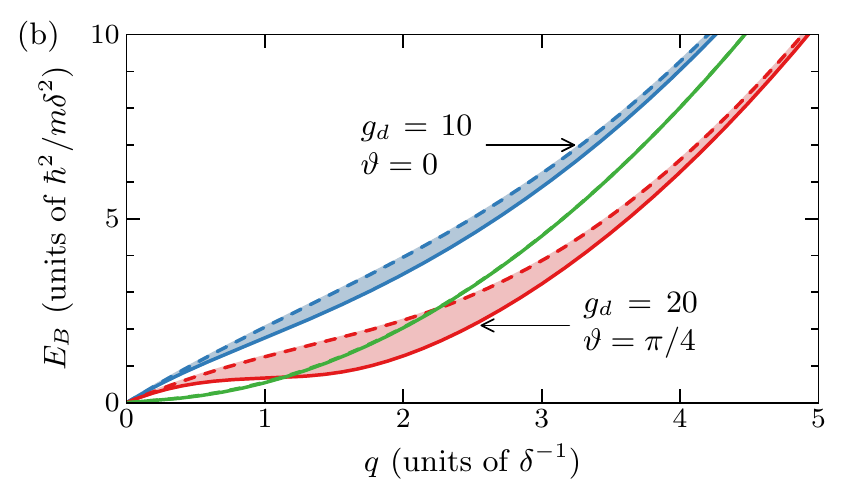}
  \caption{(Color online) Bogoliubov energies for different
    polarizations and DDI strengths.  The polar plots in (a) are
    marked with the magnitude and angle of $\bv q$.  White areas mark
    unstable regions.  (b): Cuts through Bogoliubov energies at the
    polar angles indicated in (a).  Solid lines include intra- and
    interlayer DDI, whereas dashed lines only include the intralayer
    DDI.  The green line represents \textsuperscript{52}Cr.  The
    interlayer DDI does not influence high energies where the in-plane
    excitations become particle-like.  Parameters are as in
    Fig.~\ref{fig:proj1d} with $g = 0$ and $\nu =
    1/10$.}\label{fig:Bogoliubov}
\end{figure}

Figure~\ref{fig:Bogoliubov} shows the Bogoliubov energy
Eq.~\eqref{eq:E_B} of a dipole-dominated quasi-2D multilayer BEC for
three polarization directions.  For nonperpendicular polarizations the
Bogoliubov energy becomes anisotropic with higher energies along the
projected polarization direction.  In Fig.~\ref{fig:Bogoliubov} we
observe the instability at low momenta for $\vartheta = \pi/2$.  The
cuts in Fig.~\ref{fig:Bogoliubov}(b) show the development of a roton
minimum at moderately large DDI strength.  The interlayer DDI advances
the development of this minimum to smaller values of $g_d$ compared to
a single layer quasi-2D dipolar BEC.  For comparison we also plot the
Bogoliubov energy for \textsuperscript{52}Cr in
Fig.~\ref{fig:Bogoliubov}(b), where we assumed that the contact
interaction has been reduced to $g = 0$ via a Feshbach
resonance~\cite{KocLahMet08}.  The interlayer DDI strength of
\textsuperscript{52}Cr is too weak to influence the dispersion
significantly.

\section{Conclusion}\label{sec:conclusion}
We showed that interlayer DDI in a multilayer stack of dipolar BECs
markedly reduces the aspect ratio of the quasi-2D BEC in the central
layer.  The greatest change in aspect ratio occurs for parallel
polarization.  We suggested that this effect of the interlayer DDI is
observable in time of flight image of the central layer.

To simplify numerical computations we presented a 2D model for a stack
of quasi-2D dipolar BECs created by a strong 1D optical lattice and
transversely trapped in a harmonic potential.  Our model is based on
a dimension reduction of the GPE assuming a Gaussian axial density
profile of the wave function in the individual layers.  We derived
effective intra- and interlayer DDI potentials for the resulting
coupled quasi-2D BECs.  For weak interlayer DDI we observed only small
variations in the particle numbers per lattice site, which allowed us
to derive a single mode approximation for the quasi-2D BECs in the
central sites.  This approximation reduces the numerical computation
of mean-field ground states of this system from $\sim 1$ day to
several seconds.  The resulting ground states match the reduced ground
states of the 3D GPE excellently up to moderately large DDI strengths.
For large DDI strengths $g_d \simeq g$ we still found very good
agreement at all polarizations.

Finally, the interlayer DDI reduces the squared Bogoliubov energy,
which influences the development of a roton minimum and possibly leads
to an instability (imaginary energy) for large density or DDI
strength.  The excitation spectrum of local perturbations becomes
anisotropic for nonperpendicular polarization.

\begin{acknowledgments}
  We are grateful for fruitful discussions with Dieter Jaksch and Uwe
  Fischer. This work was supported by the Academic Research Fund of
  Ministry of Education of Singapore Grant No.  R-146-000-120-112.
\end{acknowledgments}

\appendix

\section{Derivation of the effective 2D model}\label{app:derivation}
In this appendix we present the derivation of the effective 2D model
for multilayered dipolar BECs in a 1D optical lattice
[Eq.~\eqref{eq:V_2d}].  First we use the identity $U_\text{dd}(\bv r)
= -c_\text{dd} [\delta(\bv r)/3 + \partial_{\bv d\bv d} (1/4\pi|\bv
r|)]$ to split the DDI into a local and nonlocal
part~\cite{ODeGioEbe04,BaoCaiWan10}.  Then we insert $\psi(\bv r, t) =
\eu^{-\im t/2\gamma^2} \sum_j \psi_j(\bv\rho, t) w_j(z)$ with $w_j(z)
= (1/\pi\gamma^2)^{1/4} \eu^{-(z-z_j)^2/2\gamma^2}$ into
Eq.~\eqref{eq:GPE}, where we approximate $V_o(z) \simeq
\tfrac{1}{2\gamma^4} \sum_j (z-z_j)^2$.  We multiply by
$w_\ell(z)$ and integrate the resulting equation over $z$.  Setting
$\int dz w_\ell(z) w_j(z) = 0$ for $\ell\neq j$ and using $\int dz
w_\ell^2(z) = 1$, $\int dz w_\ell^4(z) = 1/\sqrt{2\pi\gamma^2}$ we
find
\begin{equation}\label{eq:app:GPE1}
  \im\partial_t \psi_\ell = \left[-\frac{1}{2}\nabla_\perp^2 + V_\text{ho} +
    \bar g (1 - \epsilon_\text{dd}) |\psi_\ell|^2 \right] \psi_\ell +
    \Psi_\ell.
\end{equation}
Here, $\nabla_\perp = \partial_{xx} + \partial_{yy}$ and
\begin{equation}\label{eq:app:Phi_2d}
  \begin{split}
    \Psi_\ell &= -3g_d \int dz d^3\bv r' w_\ell(z) \partial_{\bv
      d\bv d} U_\text{3D}(\bv r - \bv r')\\
    &\quad \times \sum_{j,p,q} \psi_j^*(\bv\rho', t) \psi_p(\bv\rho',
    t) \psi_q(\bv\rho, t) w_j(z') w_p(z') w_q(z).
  \end{split}
\end{equation}
The kernel in Eq.~\eqref{eq:app:Phi_2d} fulfills $\nabla^2
U_\text{3D}(\bv r) = -\delta(\bv r)$ so that for any $f = f(\bv r)$
\begin{equation}\label{eq:app:U_3d}
  \partial_{zz} (U_\text{3D} \star f) = - f - \nabla_\perp^2
  (U_\text{3D} \star f),
\end{equation}
where $\star$ denotes a convolution.  We expand the directional
derivative in Eq.~\eqref{eq:app:Phi_2d} as $\partial_{\bv d\bv d}
= \partial_{\bv d_\perp\bv d_\perp} + d_z^2\partial_{zz} +
2d_z\partial_{\bv d_\perp z}$ with $\bv d_\perp = (d_x, d_y)$.
Applying Eq.~\eqref{eq:app:U_3d} to the convolution in
Eq.~\eqref{eq:app:Phi_2d} yields
\begin{equation}\label{eq:app:Psi_ell2}
  \begin{split}
    \Psi_\ell &= 3g_d \Biggl( \frac{d_z^2}{\sqrt{2\pi\gamma^2}} -
    \sum_{j,p,q} \int dz d^3\bv r'
    \psi_q(\bv\rho, t)  w_q(z) w_\ell(z) \\
    &\quad \times\bigl(\partial_{\bv d_\perp\bv d_\perp} -
    d_z^2\nabla_\perp^2 + 2d_z\partial_{\bv d_\perp z} \bigr)
    U_\text{3D}(\bv r - \bv r')\\
    &\quad \times\psi_j^*(\bv\rho', t) \psi_p(\bv\rho', t) w_j(z')
    w_p(z')\Biggr).
  \end{split}
\end{equation}
The first term in Eq.~\eqref{eq:app:Psi_ell2} contributes to the
contact interaction, whereas the second term forms the nonlocal
potential.

The even kernel $U_\text{even}^{j\ell}$ [Eq.~\eqref{eq:U_even}] is
determined by the terms in Eq.~\eqref{eq:app:Psi_ell2} with only
radial derivatives.  After inserting $U_\text{3D}$ and the Gaussians
$w_j$ into Eq.~\eqref{eq:app:Psi_ell2}, we need to solve the integral
\begin{equation}\label{eq:app:zz-integral}
  \iint dz dz' \frac{\eu^{-[(z' - z_j)^2 +
      (z' - z_p)^2 + (z - z_q)^2 + (z - z_\ell)^2]/2\gamma^2}}{4\pi^2\gamma^2\sqrt{(x -
      x')^2 + (y - y')^2 + (z - z')^2}}.
\end{equation}
We substitute $\zeta = z-z' - (z_q+z_\ell-z_j-z_p)/2$, $\zeta' = z+z'
- (z_q+z_\ell+z_j+z_p)/2$ in Eq.~\eqref{eq:app:zz-integral} and
integrate over $\zeta'$.  The solution defines the even kernel of the
DDI potential with $\rho = \sqrt{(x-x')^2 + (y-y')^2}$
\begin{equation}\label{eq:app:U_even}
  U_\text{even}^{jpq\ell}(\rho) =
  \frac{1}{2(2\pi)^{3/2}\gamma} \int d\zeta
  \frac{\eu^{-\zeta^2/2\gamma^2} \eu^{-\bigl(\delta_{jp}^2 +
      \delta_{q\ell}^2 \bigr)/4\gamma^2}}{\sqrt{\rho^2 + \left(\zeta +
        \frac{\delta_{qj} + \delta_{\ell p}}{2} \right)^2}}.
\end{equation}
In Fourier space with $\bv k = k(\cos\varphi, \sin\varphi)$ the
derivatives $\partial_{\bv d_\perp\bv d_\perp} - d_z^2\nabla_\perp^2$
in Eq.~\eqref{eq:app:Psi_ell2} become $-k^2[(d_x\cos\varphi +
d_y\sin\varphi)^2 -d_z^2]$.  We use the convention $\hat f(\bv k) =
(1/2\pi) \int d^2\bv\rho f(\bv\rho) \eu^{-\im \bv k\cdot\bv\rho}$ for
the 2D Fourier transform.  With this normalization the convolution
theorem is $\mathcal{F}[f\star h] = 2\pi\mathcal{F}[f]\mathcal{F}[h]$.
For radially symmetric $f(\bv\rho)=f(\rho)$: $\hat f(k) = \int d\rho
\rho f(\rho) J_0(k\rho)$ with $J_0$ the Bessel function.  Using this
formula for the Fourier transform of the Eq.~\eqref{eq:app:U_even} and
multiplying by $2\pi k^2$ from the convolution and the Fourier
transform of the derivatives in Eq.~\eqref{eq:app:Psi_ell2} we find
\begin{equation}\label{eq:app:U_even-hat}
  \begin{split}
    \hat U_\text{even}^{jpq\ell}(k) &= \frac{k}{4} \Biggl[
    \eta\left( \frac{\gamma^2 k + (\delta_{qj} + \delta_{\ell
          p})/2}{\sqrt{2\gamma^2}} \right)\\
    &\quad + \eta\left( \frac{\gamma^2 k - (\delta_{qj} + \delta_{\ell
          p})/2}{\sqrt{2\gamma^2}} \right)
    \Biggr]\\
    &\quad \times \eu^{-\tfrac{2\delta_{jp}^2 + 2\delta_{q\ell}^2 +
        (\delta_{qj}+\delta_{\ell p})^2}{8\gamma^2}}.
  \end{split}
\end{equation}
For $j=p=q=\ell$ Eq.~\eqref{eq:app:U_even-hat} reduces to the
intralayer DDI $\hat U_\text{2D}(k)$.  Because of the exponential
prefactor, terms where all $j, p, q, \ell$ are mutually unequal are
strongly suppressed.  Similarly, terms with $q = j$, $p = \ell$ and
$j\neq\ell$ are exponentially suppressed.  The remaining terms $q =
\ell$, $p = j$, and $j\neq\ell$ form the interlayer DDI kernel $\hat
U_\text{even}^{j\ell}$ [Eq.~\eqref{eq:U_even}].

The odd kernel $U_\text{odd}^{j\ell}$ [Eq.~\eqref{eq:U_odd}] is
determined by the term in Eq.~\eqref{eq:app:Psi_ell2} with derivative
$\partial_{\bv d_\perp z}$.  Using $\partial_z(U_\text{3D} \star g) =
(\partial_z U_\text{3D}) \star g$ we insert the derivative $\partial_z
U_\text{3D}$ into Eq.~\eqref{eq:app:Psi_ell2}.  Then we need to solve
the integral
\begin{equation}
  -\iint dz dz' \frac{(z-z')\eu^{-[(z' - z_j)^2 +
      (z' - z_p)^2 + (z - z_q)^2 + (z - z_\ell)^2]/2\gamma^2}}{4\pi^2\gamma^2 [(x -
    x')^2 + (y - y')^2 + (z - z')^2]^{3/2}}.
\end{equation}
Following the steps for the even kernel we obtain the odd kernel
\begin{equation}\label{eq:app:U_odd}
  \begin{split}
    U_\text{odd}^{jpq\ell}(\rho) &= -\frac{1}{2(2\pi)^{3/2}\gamma}
    \int d\zeta \left( \zeta + \frac{\delta_{qj} + \delta_{\ell p}}{2}
    \right)\\
    &\quad \times\frac{\eu^{-\zeta^2/2\gamma^2}
      \eu^{-\bigl(\delta_{jp}^2 + \delta_{q\ell}^2
        \bigr)/4\gamma^2}}{\Bigl[\rho^2 + \Bigl(\zeta +
      \frac{\delta_{qj} + \delta_{\ell p}}{2} \Bigr)^2 \Bigr]^{3/2}}.
  \end{split}
\end{equation}
The Fourier transform of $U_\text{odd}^{jpq\ell}$
[Eq.~\eqref{eq:app:U_odd}] multiplied by $2\pi k$ from from the
Fourier transforms of the convolution and the remaining radial
derivative is given by
\begin{equation}\label{eq:app:U_odd-hat}
  \begin{split}
    \hat U_\text{odd}^{jpq\ell}(k) &= \frac{k}{4} \Biggl[
    \eta\left( \frac{\gamma^2 k + (\delta_{qj} + \delta_{\ell
          p})/2}{\sqrt{2\gamma^2}} \right)\\
    &\quad - \eta\left( \frac{\gamma^2 k - (\delta_{qj} + \delta_{\ell
          p})/2}{\sqrt{2\gamma^2}} \right)
    \Biggr]\\
    &\quad \times \eu^{-\tfrac{2\delta_{jp}^2 + 2\delta_{q\ell}^2 +
        (\delta_{qj}+\delta_{\ell p})^2}{8\gamma^2}}.
  \end{split}
\end{equation}
Only terms with $q=\ell$, $p=j$ are not exponentially suppressed in
Eq.~\eqref{eq:app:U_odd-hat}.  Hence, we recover $\hat
U_\text{odd}^{j\ell}$ [Eq.~\eqref{eq:U_odd}].

By combining Eqs.~\eqref{eq:app:U_even-hat}
and~\eqref{eq:app:U_odd-hat} with Eq.~\eqref{eq:app:Psi_ell2} and
neglecting the suppressed terms in the sum we recover the DDI
potential Eq.~\eqref{eq:V_2d} in Fourier space.

For completeness we present an approximation of the spatial potential
for multilayer DDI with arbitrary polarization direction.  To obtain
this approximation we take the limit $\gamma\rightarrow 0$ in
Eqs.~\eqref{eq:app:U_even} and~\eqref{eq:app:U_odd} treat the
Gaussians in $\zeta$ as approximations for the Dirac delta
distribution:
\begin{align}
  \lim_{\gamma\rightarrow 0} U_\text{even}^{j\ell}(\rho) &=
  \frac{1}{4\pi}
  \frac{1}{\left(\rho^2 + \delta_{\ell j}^2 \right)^{1/2}},\\
  \lim_{\gamma\rightarrow 0} U_\text{odd}^{j\ell}(\rho) &=
  -\frac{1}{4\pi} \frac{\delta_{\ell j}}{\left(\rho^2 + \delta_{\ell j}^2
    \right)^{3/2}}.
\end{align}
Again we neglect the exponentially suppressed terms.  Inserting these
kernels into Eq.~\eqref{eq:app:Psi_ell2} and calculating the remaining
derivatives we find
\begin{equation}\label{eq:app:V_2d}
  V_\text{2D}^\ell(\bv\rho) = 3g_d \sum_j
  \int d\bv\rho' U_\text{2D}^{j\ell}(\bv\rho - \bv\rho')
  |\psi_j(\bv\rho', t)|^2
\end{equation}
with
\begin{equation}\label{eq:app:U_2d}
  \begin{split}
    U_\text{2D}^{j\ell}(\bv\rho) &\simeq \frac{1}{4\pi \left( \rho^2 +
        \delta_{\ell j}^2
      \right)^{5/2}}\bigl[\rho^2 +(1- 3d_z^2)\delta_{\ell j}^2\\
    &\quad -6d_z\delta_{\ell j} \bv d_\perp\cdot\bv\rho - 3|\bv
    d_\perp \cdot\bv\rho|^2 \bigr].
  \end{split}
\end{equation}
For the intralayer part $j=\ell$ this approximation remains valid for
$\rho \gg \gamma$.  We note that Eq.~\eqref{eq:app:U_2d} corresponds
to the dimensionless DDI potential Eq.~\eqref{eq:U_dd} projected onto
2D planes separated by $\delta_{\ell j}$.

\bibliographystyle{apsrev4-1}
\bibliography{diplayers}
\end{document}